\documentclass[10pt]{iopart}

\usepackage{graphicx}
\usepackage{iopams}
\usepackage{enumerate}
\usepackage{bm}
\usepackage{cite}
\usepackage{tabularx}
\usepackage{tabu}
\usepackage{booktabs}

\usepackage{iopams}
\expandafter\let\csname equation*\endcsname\relax
\expandafter\let\csname endequation*\endcsname\relax
\usepackage{amsmath}
\begin{document}

\title{Stereoscopic molecular tagging for superconducting accelerator-cavity quench spot detection}
\author{Shiran Bao$^{1,2}$, Toshiaki Kanai$^{2,3}$, Yang Zhang$^{1,4}$, Louis N. Cattafesta III$^{1,4}$ and Wei Guo$^{1,2,*}$}
\address{$^1$ Department of Mechanical Engineering, FAMU-FSU College of Engineering, Florida State University, Tallahassee, Florida 32310, USA}
\address{$^2$ National High Magnetic Field Laboratory, Florida State University, 1800 E Paul Dirac Dr., Tallahassee, Florida 32310, USA}
\address{$^3$ Department of Physics, Florida State University, Tallahassee, Florida 32306, USA}
\address{$^4$ Florida Center for Advanced Aero-Propulsion, Florida State University, 2003 Levy Ave., Tallahassee, Florida 32310, USA}
\vspace{10pt}
\begin{indented}
\item[]* Corresponding: wguo@magnet.fsu.edu
\end{indented}
\vspace{10pt}
\begin{indented}
\item[]May 2020
\end{indented}

\begin{abstract}
Superconducting radio-frequency (SRF) cavities cooled by superfluid helium-4 (He II) are building blocks of many modern particle accelerators due to their high quality factor. However, Joule heating from sub-millimeter surface defects on cavities can lead to cavity quenching, which limits the maximum acceleration gradient of the accelerators. Developing a non-contacting detection technology to accurately locate these surface defects is the key to improve the performance of SRF cavities and hence the accelerators. In a recent proof-of-concept experiment (Phys. Rev. Applied, \textbf{11}, 044003 (2019)), we demonstrated that a molecular tagging velocimetry (MTV) technique based on the tracking of a He$_2^*$ molecular tracer line created nearby a surface hot spot in He II can be utilized to locate the hot spot. In order to make this technique practically useful, here we describe our further development of a stereoscopic MTV setup for tracking the tracer line's motion in three-dimensional (3D) space. We simulate a quench spot by applying a transient voltage pulse to a small heater mounted on a substrate plate. Images of the drifted tracer line, taken with two cameras from orthogonal directions, are used to reconstruct the line profile in 3D space. A new algorithm for analyzing the 3D line profile is developed, which incorporates the finite size effect of the heater. We show that the center location of the heater can be reproduced on the substrate surface with an uncertainty of only a few hundred microns, thereby proving the practicability of this method.
\end{abstract}

\vspace{2pc}


%
\ioptwocol

\section{\label{sec:introduction}Introduction}
Superconducting radio-frequency (SRF) cavities are key components of modern particle accelerators due to their high quality factor and low surface resistance~\cite{padamsee_rf_2008,Padamsee-2017-SUST}. Most SRF cavities have thin-walled structures made of pure niobium and are normally cooled by immersion in a liquid helium bath, especially in superfluid helium-4 (He II) below 2.17 K. Carefully fabricated cavities can achieve an acceleration gradient of 30-35 MV/m or higher~\cite{Padamsee-2020-booksection}. However, tiny defects on the cavity inner surface (e.g. micro-particle contaminants, cracks, and scratches) can cause Joule heating during the operation of the cavities, which raises the local temperature and leads to the formation of hot spots. When the temperature of these hot spots exceeds the superconducting transition temperature (i.e., 9.3 K for niobium), cavity quenching occurs and the energy stored in the cavity quickly converts to heat at the defect location in a few milliseconds. This heat is conducted through the cavity wall, leading to a heated area of order 1 cm$^2$ on the cavity outer surface~\cite{padamsee_rf_2008}. Cavity quenching often limits the acceleration gradient to below 10-15 MV/m~\cite{Padamsee-2020-booksection}. Therefore, locating the quench spots for defect removal is critical for improving the performance of SRF cavities.

So far, two major diagnostic techniques, i.e., temperature mapping (T-mapping) and second-sound triangulation, have been developed for SRF cavity quench-spot detection~\cite{Conway2017Instrumentation,watanabe_cavity_2011}. The T-mapping method, which has been widely utilized at accelerator labs, is based on the installation of a large number of (i.e., over 10$^3$) temperature sensors on the cavity outer surface ~\cite{Knobloch1994Design,canabal_full_2008} or a few tens of sensors on a rotating arm which can scan across the cavity surface~\cite{shu_novel_1996, sawamura_cavity_2008}. When the cavity is energized to just below the quench threshold, a hot spot can be identified as a location with a slightly higher temperature~\cite{Conway2017Instrumentation}. However, the drawback of the T-mapping method is clear: for fixed systems, it is extremely time and labor consuming to install and calibrate the large amount of sensors; while for scanning systems, the detection sensitivity is typically low due to the gap between the sensors and the cavity surface required for smooth rotation of the arm~\cite{Conway2017Instrumentation}.

The second-sound triangulation method is a non-contacting quench-spot detection method developed by Conway and Hartill~\cite{conway_defect_2010}. In He II, phenomenologically there exist two interpenetrating fluid components: a viscous normal fluid that carries all the entropy and an inviscid superfluid that has zero entropy~\cite{Tilley-book}. Due to the two-fluid nature, two distinct sound-wave modes exist in He II: an ordinary pressure-density wave (first sound) where both fluids move in phase, and a temperature-entropy wave (second sound) where the two fluids move out of phase~\cite{Landau-book}. When a quench spot deposits heat in He II, second-sound waves are emitted in He II towards all directions, which can be detected using various types of sensors ~\cite{Sherlock1970Oscillating, Shepard1979Development, Lunt2017Towards}. Triangulation of the quench spot can be performed by measuring the second-sound arrival time with three or more sensors. However, the reliability of this method is questioned by a puzzling observation that the triangulation does not converge on the cavity surface unless a second-sound velocity $c_2$ higher than literature values is assumed~\cite{Padamsee-2020-booksection,bertucci_quench_2013}. Despite extensive research on this puzzle~\cite{peters_advanced_2014,markham_quench_2015, eichhorn_mystery_2014,junginger_high_2015, Plouin-2019-PRAB, Bao-2019-PRApplied}, there still lacks a consensus, and the spatial resolution of the triangulation method is limited to about 1 cm~\cite{Padamsee-2017-SUST}.

In our recent proof-of-concept experiment~\cite{Bao-2019-PRApplied}, we demonstrated a novel non-contacting quench-spot detection method based on molecular tagging velocimetry (MTV) in He II. A miniature heater mounted on a substrate plate in He II was pulsed on to simulate a surface quench spot. A thin line of He$_2^*$ molecular tracers were then created in the vertical imaging plane above the heater via femtosecond laser-field ionization of helium atoms \cite{Gao2015Producing}. This tracer line deforms due to the heat-induced flow in He II~\cite{Guo2014Visualization}. By analyzing the line profile, we were able to locate the heater along the one-dimensional line where the imaging plane intersects with the substrate.

In order for this method to be practically useful, we need to demonstrate the capability of detecting a hot spot on a two-dimensional (2D) surface. For this purpose, the heater needs to be placed at an arbitrary location not in the same vertical plane as the tracer line. Therefore, the deformation of the tracer line is no longer confined to the vertical imaging plane. A stereoscopic MTV system is thus needed for capturing the three-dimensional (3D) profile of the drifted tracer line. In this paper, we report the application of such a stereoscopic MTV system for hot-spot detection on a 2D substrate surface. The experimental procedures are described in Sec.~\ref{sec:exp}. A new algorithm for analyzing the drifted tracer line in 3D space is discussed in Sec.~\ref{sec:fitmodel}. We present the analysis results in Sec.~\ref{sec:result} to show that the center of the heater can be located with an unprecedented resolution of a few hundred microns. A brief summary is given in Sec.~\ref{sec:summary}.

\section{\label{sec:exp} Experimental procedures}
\begin{figure}[h]
\centering
\includegraphics[width=0.95\columnwidth]{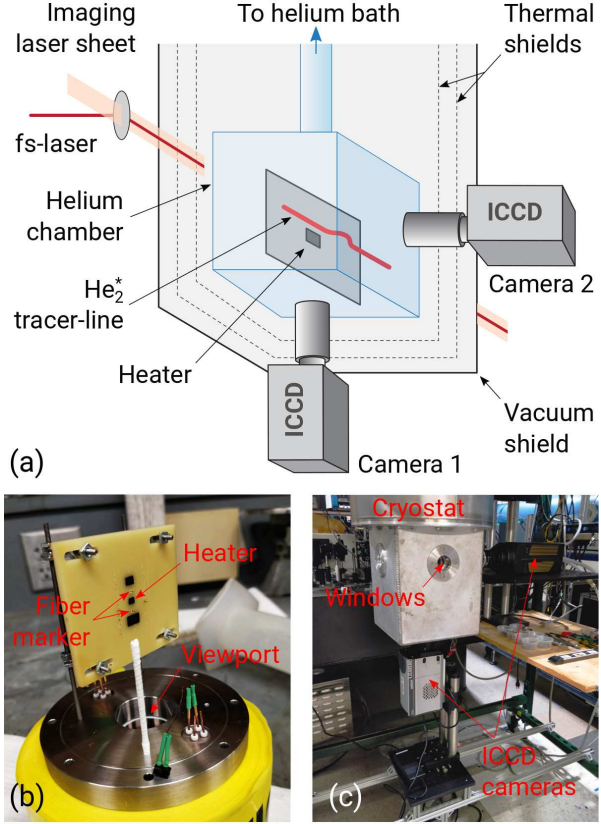}
\caption{(a) Schematic diagram of the stereoscopic MTV experimental setup for hot-spot detection; (b) A photo of the heater and the substrate assembly; (c) A photo of the cryostat and the ICCD cameras.}
\label{Fig1}
\end{figure}

A schematic diagram of the stereoscopic MTV system for hot-spot detection in He II is shown Fig.~\ref{Fig1}~(a). An aluminum cubic helium chamber (inner side width: 3 inches) with four optical windows is connected to a liquid helium bath inside an optical cryostat. The temperature of the helium in the bath, controlled by regulating the bath vapor pressure, is maintained at 1.85 K, typical of the operation temperature of SRF cavities. The helium chamber and the bath are all shielded by multi-layer vacuum insulation blankets and a liquid nitrogen thermal shield. A 50-$\Omega$  metal-foil resistor heater (surface area $A_h$=3.5$\times$2.5 mm$^2$) is mounted on the surface of a G-10 substrate plate (see Fig.~\ref{Fig1}~(b)) to simulate a surface hot spot. The plate is installed vertically inside the helium chamber at a location that can be conveniently observed from both the side and the bottom windows of the cryostat. We also drilled a few holes in the G10 plate and placed four optical fibers through them to serve as the reference points for length-scale calibration of the images.

When a voltage pulse with a duration $\Delta t$ is applied to the heater, some heat deposited in He II is carried away by a second-sound zone propagating at the speed $c_2$=19.5 m/s at 1.85 K. Inside this second-sound zone, a thermal counterflow establishes spontaneously~\cite{VanSciver2012Helium}. The normal fluid flows away from the heat source at a velocity given by $v_n$=$q/\rho sT$, where $q$ is the local heat flux, $\rho$ and $s$ are the helium density and specific entropy, respectively; and the superfluid moves in the opposite direction to compensate the fluid mass~\cite{Landau-book}. As revealed in our previous study ~\cite{Bao-2019-PRApplied}, this counterflow velocity field contains important information such as the location of the heat source and the actual heat energy carried by the second-sound waves. To probe this normal-fluid velocity field, we implement the MTV technique developed in our lab by focusing a 5-kHz femtosecond (fs) laser beam (wavelength about 800 nm) through the helium chamber~\cite{Gao2015Producing}. This fs-laser beam can create a thin line of He$_2^*$ molecular tracers with a 13-s radiative decay lifetime~\cite{McKinsey-PRA_1999}. Due to their small size (about 6 {\AA} in radius~\cite{Guo-2020-PRB}), these molecules are entrained by the viscous normal fluid above 1 K without being affected by the superfluid vortices~\cite{Zmeev-2013-PRL}, thereby allowing quantitative normal-fluid velocity-field measurements~\cite{Marakov2015Visualization, Gao-PRB_2016, Gao-JETP_2016, Gao-PRB_2017, Gao-2017-JLTP, Gao-2018-PRB, Varga-PRF_2018, Bao-2018-PRB}. We then utilize a laser-induced fluorescence technique to image the tracer line~\cite{Guo-2009-PRL}. A 905-nm pulsed laser is shaped into a 1-mm thick laser sheet that covers the entire region traversed by the tracer line. This imaging laser drives the He$_2^*$ molecules to an excited electronic state, the decay from which leads to the emission of 640-nm fluorescent light. To enhance the fluorescence efficiency, continuous fiber lasers at 1073 nm and 1099 nm are also used to recover the molecules lost to unwanted vibrational levels \cite{Guo-2009-PRL,Guo-2010-JLTP,Guo-2010-PRL}. The fluorescent light is then captured by two intensified CCD (ICCD) cameras synchronized with the imaging laser from the side and the bottom windows, as shown in Fig.~\ref{Fig1}~(c).

\begin{figure} [b]
	\centering
	\includegraphics[width=0.95\columnwidth]{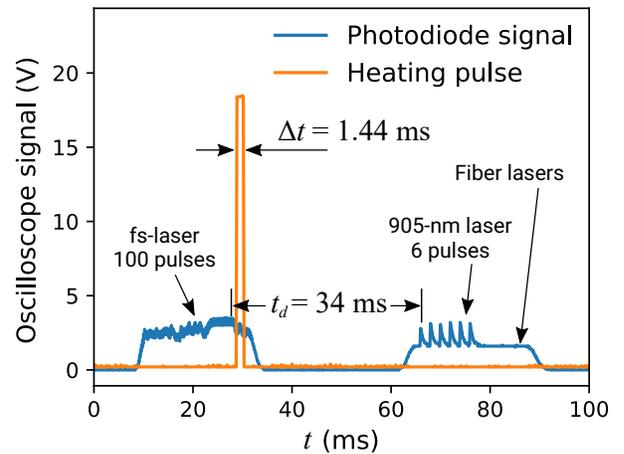}
	\caption{Oscilloscope signals from the photodiode and the heater drive showing the typical timing setting of the experiment.}
	\label{Fig2}
\end{figure}

In our experiment, we first send in 100 fs-laser pulses to create a clear He$_2^*$ tracer line. Then, a rectangular voltage pulse with a duration $\Delta t$=1.44 ms and an adjustable magnitude $V_0$ up to 36.8 V is applied to the heater. After a drift time $t_d$, 6 pulses from the 905-nm imaging laser at a repetition rate of 500 Hz are used to illuminate the tracer line for imaging purpose. Normally, a baseline image is taken at zero drift time as a reference. Then, we repeat the measurements 5 times at a sufficient drift time (i.e., $t_d$=34 ms) such that the deformation of the tracer line due to the entire second-sound zone can be captured. Fig.~\ref{Fig2} shows the typical time sequence of the laser pulses recorded by a photo-diode sensor near the entrance window of the cryostat together with the heat pulse.

\section{\label{sec:fitmodel} Tracer-line deformation and analysis model}
In our previous study~\cite{Bao-2019-PRApplied}, a miniature heater was used, and the analysis of the drifted tracer-line profile was performed assuming isotropic heat transfer from a point heat source, which greatly simplified the trial-and-error procedure for determining the heater location. However, in the current experiment the heater size is increased by nearly nine times in order to better represent a quench spot, which makes the point heat-source model inapplicable. Therefore, a new analysis model that accounts for the finite size effect of the heater must be implemented.

\begin{figure}
\centering
\includegraphics[width=0.95\columnwidth]{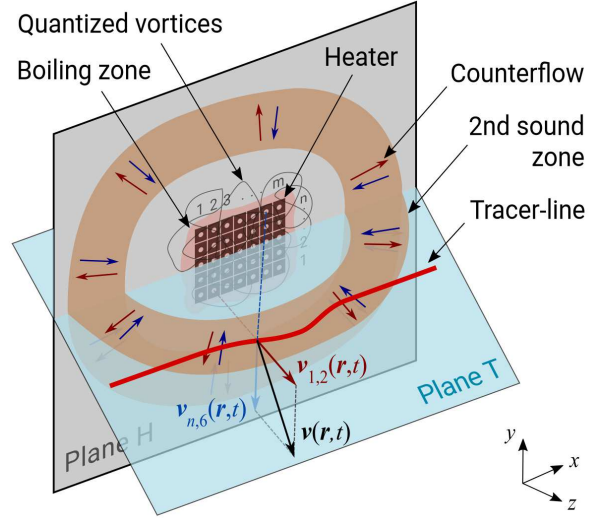}
\caption{A schematic diagram showing the concept of the method for evaluating the drifted tracer-line profile by discretizing the heater. The Plane H denotes the heater plane, and Plane T is where the drifted tracer line resides.}
\label{Fig3}
\end{figure}

To aid the discussion of this analysis model, let us first briefly outline the relevant heat transfer processes in He II. Following the application of the voltage pulse, the Joule heat generated by the heater is deposited in He II in three distinct processes. First, some heat energy is consumed in the formation of a small vapor zone which encloses the heater surface (see Fig.~\ref{Fig3}). This is because the instantaneous heat flux from the heater surface, i.e., in the range of 77 to 310 W/cm$^2$ as typical for quench spots, is well above the threshold for boiling in He II (i.e., about 15 W/cm$^2$ for a heat pulse of a few milliseconds~\cite{Shimazaki1995Second,iida_visualization_1996,Hilton2005Direct}). Then, outside this vapor zone, some heat energy is carried out by the propagating second-sound zone. The tracer line deforms as the second-sound zone passes across it, due to the associated normal-fluid flow. The third process is the creation of quantized vortices in the superfluid due to the counterflow of the two fluids following the second-sound wavefront~\cite{Vinen1957Mutual}. However, as we already revealed~\cite{Bao-2019-PRApplied}, dense vortices only emerge within a thin layer of He II outside the vapor zone, because the heat flux drops rapidly away from the heater surface in 3D space. The thermal energy stored in this layer is small, and it spreads out diffusively, which hardly affects the tracer line. As the heat pulse ends, the vapor zone collapses. The released thermal energy is carried out largely by the emission of first-sound shock waves. But since the first sound only causes the fluid parcels to oscillate, there is again little detectable effect on the tracer line.

To calculate the expected profile of the drifted tracer line, we start with an assumed center location of the heater on the substrate. In principle, a direct numerical simulation (DNS) of the full-space normal-fluid velocity field can be conducted by solving simultaneously the conservation equations of the He II mass, momentum, and energy, coupled with the evolution equation of the quantized vortices~\cite{Zhang-2006-IJHMT}. Then, the drifted-line profile can be derived based on the initial line location and the local normal-fluid velocity. By adjusting the assumed heater location to achieve the best match between the calculated and the observed drifted tracer-line profiles, the accurate location of the heater can be determined. However, a DNS-based trial-and-error method is extremely time consuming. Instead, a simplified model based on the discretization of the finite-size heater is adopted.

As shown schematically in Fig.~\ref{Fig3}, we divide the assumed heater into $n\times m$ small elements and treat each element as a standalone point heat source. The velocity $\bm{v}(\bm{r},t)$ of a tracer-line segment located at $\bm{r}$ is calculated simply as the vector sum of the normal fluid velocity $\bm{v}_{i,j}(\bm{r},t)$ produced by each heater element:
\begin{equation}
\centering
\bm{v}(\bm{r},t)=\sum_{i=1}^{n}\sum_{j=1}^{m} \bm{v}_{i,j}(\bm{r},t)=\sum_{i=1}^{n}\sum_{j=1}^{m} \frac{\bm{q}_{i,j}(t)}{\rho sT},
\label{eq:vn}
\end{equation}
where the heat flux vector $\bm{q}_{i,j}(t)$ points from the heater element at location $\bm{r}_{i,j}$ towards the tracer-line segment at $\bm{r}$, and its magnitude is given by $q_{i,j}(t)=\dot{Q}_{i,j}(t)/2\pi R_{i,j}^2$. Here $R_{i,j}=|\bm{r}_{i,j}-\bm{r}|$ is the distance between the heater element and the tracer-line segment, and $\dot{Q}_{i,j}(t)$ denotes the corresponding rate of heat transfer which is given by:
\begin{equation}
\dot{Q}_{i,j}(t)=\left\{
\begin{array}{cl}
0                           &   {c_2t<R_{i,j}}\\
\frac{Q_s/\Delta t}{n\times m}     &   {R_{i,j}\le c_2t \le R_{i,j}+c_2\Delta t}\\
0                           &   {c_2t > R_{i,j}+c_2\Delta t}\\
\end{array} \right.
\label{eq:effq}
\end{equation}
where $Q_s$ is the total heat energy carried away by the second-sound zone. This model assumes that the normal-fluid velocities produced by the heater elements are additive, which is reasonable when the nonlinear effects such as the mutual friction in the equation of motion of the normal fluid can be ignored. The form of $\dot{Q}_{i,j}(t)$ as given by Eq.~(\ref{eq:effq}) also ensures that the anisotropic nature of the heat transfer in 3D space from the finite-size heater can be fully captured.

In a time step $dt$, a tracer-line segment drifts by $\bm{v}(\bm{r},t)dt$. We then calculate its velocity at the new location to evaluate the displacement in the next time step. The total displacement can be obtained as $\Delta \bm{r}=\sum_i\bm{v}(\bm{r}(t_i),t_i)dt$. Repeating this calculation for every line segments, the final drifted tracer-line profile can be determined. The accuracy of this analysis depends on the total number of heater elements and the time step $dt$. Using $m$=7, $n$=5, and $dt$=0.01 ms, we find that the trial-and-error analysis converges reliably within about 10 minutes with a quad-core processor performing around 14 gigaFLOPs.

\section{\label{sec:result}Experimental results}
\subsection{Reconstruction of the 3D tracer-line profile}
\begin{figure}[t]
\centering
\includegraphics[width=0.95\columnwidth]{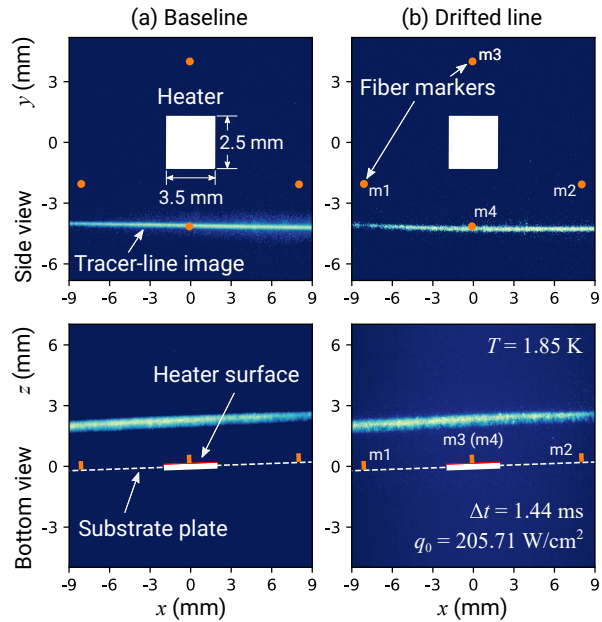}
\caption{Typical side view and bottom view images showing (a) a baseline created at about 4.7 mm away from the heater center, and (b) a drifted tracer line following a heat pulse of $q_0$=205.71 W/cm$^2$ and $\Delta t$=1.44 ms.}
\label{Fig4}
\end{figure}

In our experiments, we need to first reconstruct the 3D profiles of the tracer lines so that the tracer-line deformation analysis can be performed. Fig.~\ref{Fig4} shows typical fluorescence images of the baseline and the drifted line taken by the two cameras when a voltage pulse of 30 V is applied to the heater (corresponding heat flux: 205.71 W/cm$^2$). These tracer-line images are averaged based on five snapshot images for improved visibility. The positions of the heater and the fiber markers are also indicated in these images. To reconstruct the 3D profiles of the tracer lines based on these 2D images, we first identify the pixel coordinates of the four fiber markers (i.e., m1-m4) in the side view and the bottom view images. The physical dimension in the pixel coordinates can be calibrated based on the known distances between the markers. Since the two cameras are carefully installed in orthogonal directions, the side view image provides us the $x$-$y$ coordinates of the markers while the bottom view image gives the $x$-$z$ coordinates. We then shift the two images to ensure that the $x$-coordinates of the markers in both images are identical. For convenience, we set the origin of the coordinate system to be at the center of the actual heater. Through these procedures, the absolute coordinates of an object in 3D space can be easily extracted from the side view and the bottom view images. It should be noted that our calibration is based on the plane of the markers. For the tracer line which is not in the markers' plane, there are small (i.e., $<3\%$) offsets in its 3D coordinates due to the
out-of-plane projection. In the future, a more sophisticated photogrammetric method based on the direct linear transformation for camera calibration~\cite{liu-2000_AIAAJ} can be utilized to avoid such offsets.

\begin{figure}[t]
\centering
\includegraphics[scale=0.8]{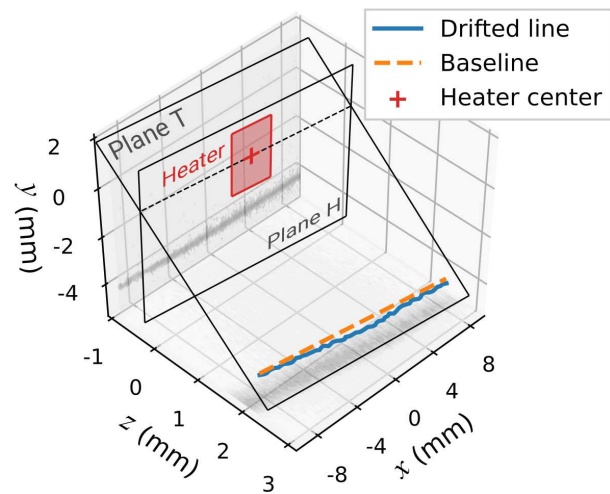}
\caption{Schematics showing the reconstructed centerline profiles in 3D space for the example baseline and drifted tracer-line images included in Fig.~\ref{Fig4}.}
\label{Fig5}
\end{figure}

We then adopt an ridge-finding algorithm~\cite{pulkkinen-2014-COA} to identify the centerlines of the fluorescent patterns in both the side view and the bottom view images of the tracer line. These centerlines are the projections of the tracer line on the $x$-$y$ plane and the $x$-$z$ plane, and therefore they allow us to easily reconstruction the 3D profile of the tracer line. Fig.~\ref{Fig5} shows the obtained 3D centerline profiles for the example baseline and drifted line included in Fig.~\ref{Fig4}. Compared to the baseline, the drifted line is slightly deformed in 3D space due to the transient heat transfer in He II. We also determine the Plane H where the heater is located through a plane fit of the markers positions. This information is needed because we will move the assumed heater in this Plane H to determine its optimum location through the analysis model discussed in Sec.~\ref{sec:fitmodel}.

\subsection{Analysis results and discussions}
\begin{figure}[tbp]
\centering
\includegraphics[width=0.95\columnwidth]{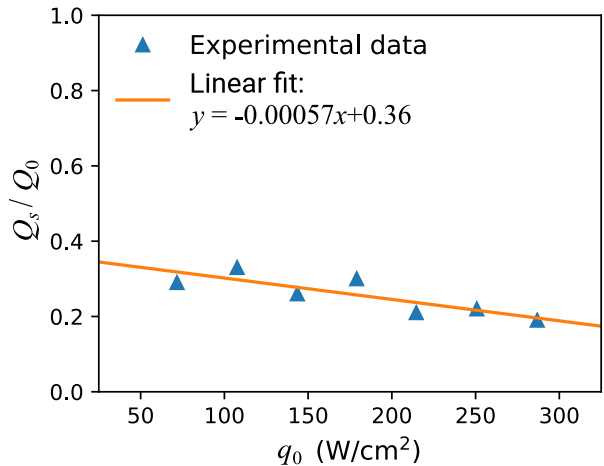}
\caption{The ratio of the heat energy $Q_s$ carried by the second-sound zone to the total heat generated by the heater $Q_0$ as a function of the heat flux $q_0$ at the heater surface. The data are taken from Ref.~\cite{Bao-2019-PRApplied} at 1.85 K.}
\label{Fig6}
\end{figure}

To perform the tracer-line deformation analysis based on the obtained 3D tracer-line profile, we need to set the values for four parameters, i.e., the coordinates of the assumed heater center location ($x_0$, $y_0$, $z_0$) and the heat energy $Q_s$ carried by the second-sound zone. Note that since the assumed heater is confined in the Plane H, only two coordinates among $x_0$, $y_0$, and $z_0$ need to be varied in the search for the best match between the calculated and the observed drifted tracer-line profiles. This situation holds also for quench-spot detection on a real SRF cavity, since the quench spot is confined in the known cavity surface. As for $Q_s$, in principle it should be treated as an unknown parameter since it is a fraction of the total heat energy $Q_0$ produced by the heater. However, in our previous study~\cite{Bao-2019-PRApplied}, we have measured the ratio $Q_s/Q_0$ and found that it scales linearly with the heat flux from the heater surface $q_0$=$Q_0/A_h\Delta t$, as shown in Fig.~\ref{Fig6}. This observation makes it possible to simplify our current analysis. Through a linear fit to the data shown in Fig.~\ref{Fig6}, the value of $Q_s$ for a given $Q_0$ can be easily determined if we assume that this scaling holds for our current experiments.

\begin{figure}[tbp]
\centering
\includegraphics[width=1\columnwidth]{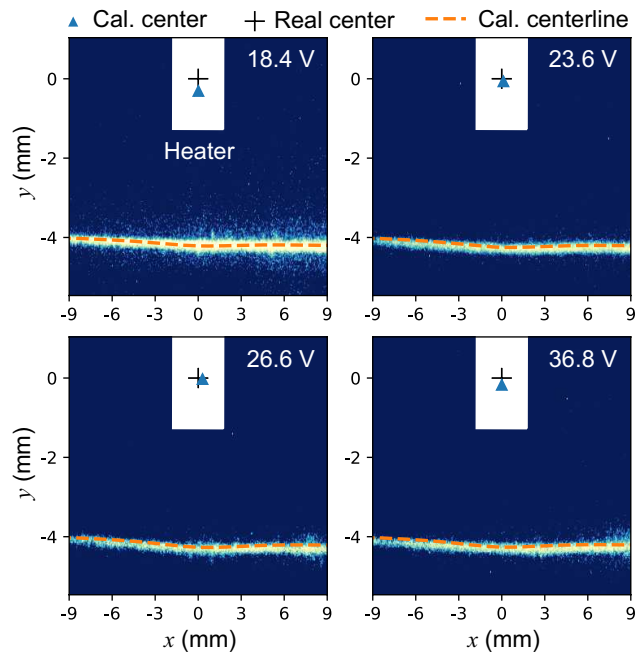}
\caption{Analysis results showing the calculated centerlines of the tracer lines projected to the $x$-$y$ plane (side view) in comparison with the actual images. The calculated and the actual heater center locations are also indicated.}
\label{Fig7}
\end{figure}

We have conducted the measurements at various voltages applied to the heater (i.e., 18.3 V to 36.8 V) and performed the tracer-line deformation analysis for all the acquired image data. Representative side-view images of the drifted tracer lines are shown in Fig.~\ref{Fig7}. The calculated drifted lines which give the least squares errors to the image data are also included. It is clear that the calculated drifted lines overlap well with the actual tracer-line images. As the voltage (and hence the heat flux) increases, the drifted tracer line exhibits a more pronounced deformation. The optimum center locations of the heater obtained through the analysis are close to the actual heater center.

\begin{table*}
	\caption{\label{tab:tab1}The calculated center location $(x_h,y_h,z_h)$ of the heater averaged over five measurements at each applied surface heat flux. The standard deviations associated with these measurements are also included.}
	\begin{indented}
		\lineup
		\item[]
		\begin{tabu} to \textwidth {X[c]X[c]|X[c]X[c]X[c]X[c]|X[c]X[c]X[c]}
			\br
			$U$&$q_0$&$x_h$&$y_h$&$z_h$&$r_h$&$\sigma_x$&$\sigma_y$&$\sigma_z$\cr
			(V)&(W/cm$^2$)&(mm)&(mm)&(mm)&(mm)&(mm)&(mm)&(mm)\cr
			\mr
			18.4&77.39&0.02&0.31&0.004&0.31&0.60&0.43&0.01\cr
			23.6&127.31&0.12&0.07&-0.001&0.13&0.56&0.33&0.02\cr
			26.6&161.73&0.30&0.03&-0.006&0.30&0.62&0.30&0.02\cr
			30.0&205.71&0.05&0.01&0.000&0.05&0.52&0.28&0.01\cr
			36.8&309.54&0.00&0.17&0.003&0.17&1.09&0.34&0.03\cr
			\br
		\end{tabu}
	\end{indented}
\end{table*}

\begin{figure}[tbp]
	\centering
	\includegraphics[width=0.95\columnwidth]{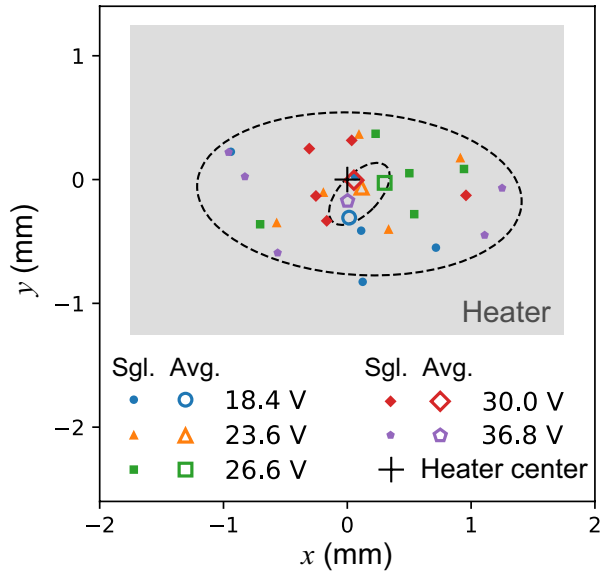}
	\caption{The calculated heater center locations based on the analysis of snapshot images (solid symbols) and five-shot averaged images (hollow symbols) at various applied voltages to the heater. The two dashed ellipses represent the 95\% confidence ellipses~\cite{tamhane-2012_book} of the solid symbols and the hollow symbols.}
	\label{Fig8}
\end{figure}

To better illustrate the accuracy of the analysis results, in Fig.~\ref{Fig8} we show the calculated heater center locations based on the analysis of individual snapshot images obtained at various applied voltages. These calculated center locations randomly distribute on the heater surface but are all within 1 mm from the actual heater center, and the overall standard deviation is 0.31 mm. We have also performed the analysis based on a superposition of 5 snapshot images at every applied voltages. This averaging can significantly improve the signal-to-noise ratio of the tracer-line profile, and therefore the obtained heat center locations appear to be notably closer to the actual heat center, i.e., with a standard deviation of only about 0.1 mm. These results together with the corresponding experimental conditions are collected in Table~\ref{tab:tab1}.

Our analysis nicely reproduces the heater center with an uncertainty of a few hundred microns, which is only a few percent of the heater size, regardless of the applied heat fluxes. This success clearly demonstrates the effectiveness and the accuracy of the stereoscopic MTV method for hot-spot detection on a 2D surface in He II. Compared to the T-mapping and the second-sound triangulation methods, the accuracy of our MTV-based method is far superior. In the future, by creating multiple tracer lines or even a tracer-line grid~\cite{Bao-2019-PRApplied,Sanavandi-2020}, it is possible to further improve the accuracy and effectiveness of the MTV method. We would also like to comment that besides the MTV technique, a particle tracking velocimetry (PTV) based flow visualization technique using frozen hydrogen or deuterium tracer particles \cite{Mastracci-2017-JLTP,Mastracci-2018-PRF,Mastracci-2019-PRF,Mastracci-2019-PRF-2} may work as well for hot-spot detection in He II, since the underlying heat transfer processes and the tracer dynamics are similar. The PTV-based method could be more easily adopted by the accelerator labs since it does not require complex laser systems and the fluorescence imaging system. We plan to conduct relevant testings and will report the result in due course.
	
\section{\label{sec:summary}Summary}	
We have conducted a carefully designed experiment to demonstrate the feasibility of a stereoscopic MTV method for surface hot-spot detection in He II. A He$_2^*$ molecular tracer line is created via fs-laser field ionization near a planar heater mounted on a substrate surface. Fluorescence images of the drifted tracer line, captured by two ICCD cameras from orthogonal directions following a heat pulse, are utilized to reconstruct its 3D profile. A simplified heat-transfer model based on a discretization of the finite-size heater is used to analyze the drifted tracer-line profile. The derived center location of the heater appears to be always within a few hundred microns from the actual heater center in the whole range of heat fluxes we explored. These results clearly prove the feasibility and accuracy of the stereoscopic MTV method for hot-spot detection, making it a promising technology for future quench-spot detection on real SRF cavities in He II.

\ack
The authors would like to acknowledge the support from the U.S. Army Research Office under Contract No. W911NF1910047. S. B. and W. G. also acknowledge the support from the U.S. Department of Energy under Grant No. {DE-SC0020113}. The experiment was conducted at the National High Magnetic Field Laboratory, which is supported by National Science Foundation Cooperative Agreement No. DMR-1644779 and the State of Florida.

\section*{References}
\bibliographystyle{iopart-num}
\bibliography{ref-extracts}

\providecommand{\newblock}{}
\begin{thebibliography}{10}
\expandafter\ifx\csname url\endcsname\relax
  \def\url#1{{\tt #1}}\fi
\expandafter\ifx\csname urlprefix\endcsname\relax\def\urlprefix{URL }\fi
\providecommand{\eprint}[2][]{\url{#2}}

\bibitem{padamsee_rf_2008}
Padamsee H, Knobloch J and Hays T 2008 {\em {{RF}} Superconductivity for
  Accelerators\/} Wiley series in beam physics and accelerator technology (New
  York: {Wiley})

\bibitem{Padamsee-2017-SUST}
Padamsee H 2017 {\em Supercond. Sci. Technol.\/} {\bf 30} 053003

\bibitem{Padamsee-2020-booksection}
Padamsee H 2020 History of gradient advances in {{SRF}} (\textit{Preprint}
  \eprint{arXiv:2004.06720})

\bibitem{Conway2017Instrumentation}
Conway Z~A, Ge M and Iwashita Y 2017 {\em Supercond. Sci. Technol.\/} {\bf 30}
  034002

\bibitem{watanabe_cavity_2011}
Watanabe K, Hayano H and Iwashita Y 2011 Cavity inspection and repair
  techniques {\em Proceedings of {{SRF2011}}\/} (Chicago) pp 598--602

\bibitem{Knobloch1994Design}
Knobloch J, Muller H and Padamsee H 1994 {\em Rev. Sci. Instrum.\/} {\bf 65}
  3521--3527

\bibitem{canabal_full_2008}
Canabal A, Tajima T, Krawczyk F, Haynes W, Roybal R, Sedillo J and Cohen S 2008
  Full real-time temperature mapping system for 9-cell {{ILC}}-type cavities
  {\em Proceedings of {{EPAC08}}\/} (Genoa) pp 841--843

\bibitem{shu_novel_1996}
Shu Q~S, Junquera T, Caruette A, Deppe G, Fouaidy M, Moeller W~D, Pekeler M,
  Proch D, Renken D and Stolzenburg C 1996 A novel rotating temperature and
  radiation mapping system in superfluid {{He}} and its successful diagnostics
  {\em Advances in {{Cryogenic Engineering}}\/} A Cryogenic Engineering
  Conference Publication (Boston: {Springer}) pp 895--904

\bibitem{sawamura_cavity_2008}
Sakai H, Shinoe K, Furuya T, Takahashi T, Umemori K and Sawamura M 2008 Cavity
  diagnostics using rotating mapping system for {{L}}-band {{ERL}}
  superconducting cavity {\em Proceedings of {{EPAC08}}\/} (Genoa) pp 907--909

\bibitem{conway_defect_2010}
Conway Z~A, Hartill D~L, Padamsee H~S and Smith E~N 2009 Defect location in
  superconducting cavities cooled with {{He}}-{{II}} using oscillating
  superleak transducers {\em Proceedings of {{SRF2009}}\/} (Berlin) pp 113--116

\bibitem{Tilley-book}
Tilley D and Tilley J 1990 {\em Superfluidity and Superconductivity\/} 3rd ed
  (Bristol: Institute of Physics)

\bibitem{Landau-book}
Landau L~D and Lifshitz E~M 1987 {\em Fluid Mechanics\/} 2nd ed vol~6 (Oxford:
  {Pergamon Press})

\bibitem{Sherlock1970Oscillating}
Sherlock R~A and Edwards D~O 1970 {\em Rev. Sci. Instrum.\/} {\bf 41}
  1603--1609

\bibitem{Shepard1979Development}
Shepard K~W, Scheibelhut C~H, Markovich P, Benaroya R and Bollinger L~M 1979
  {\em IEEE Trans. Magn.\/} {\bf 15} 666--669

\bibitem{Lunt2017Towards}
Lunt A, Kov$\rm\acute{a}$cs Z, Furci H, Koettig T, L$\rm\acute{e}$aux F and
  Vandoni G 2017 {\em Mater. Des.\/} {\bf 122} 403--404

\bibitem{bertucci_quench_2013}
Bertucci M, Bosotti A, Garolfi L, Michelato P, Monaco L, Sertore D and Pagani C
  2013 Quench detection diagnostics on 3.9 {{GHz XFEL}} cavities {\em
  Proceedings of {{SRF2013}}\/} (Paris) pp 710--713

\bibitem{peters_advanced_2014}
Peters B~J 2014 {\em Advanced Heat Transfer Studies in Superfluid Helium for
  Large-Scale High-Yield Production of Superconducting Radio Frequency
  Cavities\/} Diploma {{Thesis}} Karlsruhe Institute of Technology Karlsruhe

\bibitem{markham_quench_2015}
Eichhorn R, Hartill D, Hoffstaetter G and Markham S 2015 On quench propagation,
  quench detection and second sound in {{SRF}} cavities {\em Proceedings of
  {{SRF2015}}\/} (Whistler) pp 804--809

\bibitem{eichhorn_mystery_2014}
Eichhorn R and Markham S 2015 {\em Phys. Procedia\/} {\bf 67} 822--827

\bibitem{junginger_high_2015}
Junginger T, Horn P, Koettig T, Liao K, Macpherson A and Peters B~J 2015 High
  flux three dimensional heat transport in superfluid helium and its
  application to a trilateration algorithm for quench localization with
  {{OSTs}} {\em Proceedings of {{SRF2015}}\/} (Whistler) pp 201--204

\bibitem{Plouin-2019-PRAB}
Plouin J, Baudouy B, Four A, Charrier J~P, Maurice L, Novo J, Peters B~J and
  Liao K 2019 {\em Phys. Rev. Accel. Beams\/} {\bf 22} 083202

\bibitem{Bao-2019-PRApplied}
Bao S and Guo W 2019 {\em Phys. Rev. Applied\/} {\bf 11} 044003

\bibitem{Gao2015Producing}
Gao J, Marakov A, Guo W, Pawlowski B~T, Van~Sciver S~W, Ihas G~G, McKinsey D~N
  and Vinen W~F 2015 {\em Rev. Sci. Instrum.\/} {\bf 86} 093904

\bibitem{Guo2014Visualization}
Guo W, La~Mantia M, Lathrop D~P and Van~Sciver S~W 2014 {\em Proc. Natl. Acad.
  Sci. U. S. A.\/} {\bf 111} 4653--4658

\bibitem{VanSciver2012Helium}
Van~Sciver S~W 2012 {\em Helium Cryogenics\/} 2nd ed International cryogenics
  monograph series (New York: {Springer})

\bibitem{McKinsey-PRA_1999}
McKinsey D~N, Brome C~R, Butterworth J~S, Dzhosyuk S~N, Huffman P~R, Mattoni
  C~E~H, Doyle J~M, Golub R and Habicht K 1999 {\em Phys. Rev. A\/} {\bf 59}
  200--204

\bibitem{Guo-2020-PRB}
Guo W and Golov A~I 2020 {\em Phys. Rev. B\/} {\bf 101}(6) 064515

\bibitem{Zmeev-2013-PRL}
Zmeev D~E, Pakpour F, Walmsley P~M, Golov A~I, Guo W, McKinsey D~N, Ihas G~G,
  McClintock P~V~E, Fisher S~N and Vinen W~F 2013 {\em Phys. Rev. Lett.\/} {\bf
  110}(17) 175303

\bibitem{Marakov2015Visualization}
Marakov A, Gao J, Guo W, Van~Sciver S~W, Ihas G~G, McKinsey D~N and Vinen W~F
  2015 {\em Phys. Rev. B\/} {\bf 91} 094503

\bibitem{Gao-PRB_2016}
Gao J, Guo W and Vinen W~F 2016 {\em Phys. Rev. B\/} {\bf 94} 094502

\bibitem{Gao-JETP_2016}
Gao J, Guo W, L'vov V~S, Pomyalov A, Skrbek L, Varga E and Vinen W~F 2016 {\em
  JETP Lett.\/} {\bf 103} 648--652

\bibitem{Gao-PRB_2017}
Gao J, Varga E, Guo W and Vinen W~F 2017 {\em Phys. Rev. B\/} {\bf 96} 094511

\bibitem{Gao-2017-JLTP}
Gao J, Varga E, Guo W and Vinen W~F 2017 {\em J. Low Temp. Phys.\/} {\bf 187}
  490

\bibitem{Gao-2018-PRB}
Gao J, Guo W, Yui S, Tsubota M and Vinen W~F 2018 {\em Phys. Rev. B\/} {\bf 97}
  184518

\bibitem{Varga-PRF_2018}
Varga E, Gao J, Guo W and Skrbek L 2018 {\em Phys. Rev. Fluids\/} {\bf 3}
  094601

\bibitem{Bao-2018-PRB}
Bao S, Guo W, L'vov V~S and Pomyalov A 2018 {\em Phys. Rev. B\/} {\bf 98}
  174509

\bibitem{Guo-2009-PRL}
Guo W, Wright J~D, Cahn S~B, Nikkel J~A and McKinsey D~N 2009 {\em Phys. Rev.
  Lett.\/} {\bf 102} 235301

\bibitem{Guo-2010-JLTP}
Guo W, Wright J~D, Cahn S~B, Nikkel J~A and McKinsey D~N 2010 {\em J. Low Temp.
  Phys.\/} {\bf 158} 346

\bibitem{Guo-2010-PRL}
Guo W, Cahn S~B, Nikkel J~A, Vinen W~F and McKinsey D~N 2010 {\em Phys. Rev.
  Lett.\/} {\bf 105} 045301

\bibitem{Shimazaki1995Second}
Shimazaki T, Murakami M and Iida T 1995 {\em Cryogenics\/} {\bf 35} 645--651

\bibitem{iida_visualization_1996}
Iida T, Murakami M, Shimazaki T and Nagai H 1996 {\em Cryogenics\/} {\bf 36}
  943--949

\bibitem{Hilton2005Direct}
Hilton D~K and Van~Sciver S~W 2005 {\em J. Low Temp. Phys.\/} {\bf 141} 47--82

\bibitem{Vinen1957Mutual}
Vinen W~F 1957 {\em Proc. R. Soc. A\/} {\bf 242} 493--515

\bibitem{Zhang-2006-IJHMT}
Zhang P, Murakami M and Wang R~Z 2006 {\em Int. J. Heat Mass Trans.\/} {\bf 49}
  1384--1394

\bibitem{liu-2000_AIAAJ}
Liu T, Cattafesta L~N, Radeztsky R~H and Burner A~W 2000 {\em AIAA J.\/} {\bf
  38} 964--971 ISSN 0001-1452

\bibitem{pulkkinen-2014-COA}
Pulkkinen S, M{\"a}kel{\"a} M~M and Karmitsa N 2014 {\em Comput. Optim.
  Appl.\/} {\bf 57} 129--165

\bibitem{tamhane-2012_book}
Tamhane A~C 2012 {\em Statistical Analysis of Designed Experiments: Theory and
  Applications\/} Wiley Series in Probability and Statistics (Hoboken: Wiley)

\bibitem{Sanavandi-2020}
Sanavandi H, Bao S~R, Zhang Y, Keijzer R, Guo W and Cattafesta~III L~N 2020 A
  cryogenic-helium pipe flow facility with unique double-line molecular tagging
  velocimetry capability (\textit{Preprint} \eprint{arXiv:2003.07420})

\bibitem{Mastracci-2017-JLTP}
Mastracci B, Takada S and Guo W 2017 {\em J. Low Temp. Phys.\/} {\bf 187}
  446–452

\bibitem{Mastracci-2018-PRF}
Mastracci B and Guo W 2018 {\em Phys. Rev. Fluids\/} {\bf 3} 063304

\bibitem{Mastracci-2019-PRF}
Mastracci B and Guo W 2019 {\em Phys. Rev. Fluids\/} {\bf 4} 023301

\bibitem{Mastracci-2019-PRF-2}
Mastracci B, Bao S, Guo W and Vinen W~F 2019 {\em Phys. Rev. Fluids\/} {\bf 4}
  083305

\end{thebibliography}

\end{document}